\documentclass[12pt]{article}
\usepackage{epsfig}
\usepackage{graphicx}
\newcommand{\be}{\begin{equation}}
\newcommand{\ee}{\end{equation}}
\newcommand{\ba}{\begin{eqnarray}}
\newcommand{\ea}{\end{eqnarray}}
\begin{document}
\renewcommand{\baselinestretch}{1.1}
\small\normalsize
\renewcommand{\theequation}{\arabic{section}.\arabic{equation}}
\renewcommand{\thesection}{\arabic{section}.}
\language0
                                                       
\thispagestyle{empty}

\begin{flushright}
HUB-EP-97/9\\January 1997
\end{flushright}

\vspace*{1.5cm}

\begin{center}

{\Large \bf Monopoles and deconfinement transition in SU(2) 
lattice gauge theory}

\vspace*{0.8cm}

{\bf Georg Damm $^a$ and Werner Kerler $^{a,b}$ }

\vspace*{0.3cm}

{\sl $^a$ Fachbereich Physik, Universit\"at Marburg, D-35032 Marburg, 
Germany \\
 $^b$ Institut f\"ur Physik, Humboldt-Universit\"at, D-10115 Berlin, 
Germany\hspace{2mm}} 

\end{center}

\vspace*{2.5cm}

\begin{abstract}
We investigate SU(2) lattice gauge theory in four dimensions in the maximally 
abelian projection. Studying the effects on different lattice sizes we show 
that the deconfinement transition of the fields and the percolation transition 
of the monopole currents in the three space dimensions are precisely related. 
To arrive properly at this result the uses of a mathematically sound 
characterization of the 
occurring networks of monopole currents and of an appropriate method 
of gauge fixing turn out to be crucial. In addition we investigate 
detailed features of the monopole structure in time direction.
\end{abstract}

\newpage

\section{Introduction} \setcounter{equation}{0}

\hspace{0.35cm}

For the conjecture \cite{t76,m76} that condensing magnetic monopoles 
describe quark confinement nonperturbative support first came from compact 
U(1) lattice gauge theory \cite{bmk77,dt80}. To treat the nonabelian case 't 
Hooft \cite{t81} proposed to extract the relevant abelian degrees in SU(N) 
gauge theory by a suitable gauge fixing procedure called abelian projection. 
A nonperturbative realization of this concept has become possible on the 
lattice \cite{ksw87}. Among the possible choices of such projections the 
maximally abelian projection \cite{klsw87} appears most appropriate and has
been considered in numerous papers \cite{p96}.

To confirm this picture in Monte Carlo simulations it is attractive not only 
to measure observables but to analyze the configurations and to look for the
structures of monopole currents which are characteristic for the phases and
thus are able to signal the deconfinement transition. First analyses of 
this type have already been presented some time ago \cite{bsw91,bmm92}. A
consideration of the connection of two points by currents \cite{ipp93} is
also in this spirit. Analyses of the sizes of the structures have been given in
\cite{ekms95,kms95}. More recently a quantity called ``wrapping number'' by 
the author \cite{e96} has been used in the deconfinement phase.

Monopole currents which (on lattices with periodic boundary conditions)
wrap around the torus have been typically observed in the confining phase.
However, as soon as one goes into more detail and tries to make the
characterization precise, the problem becomes obvious that one has to deal
with complicated networks of monopole currents rather than with simple loops, 
for which the topological characterization would be straightforward.
Decomposing the networks into loops, apart from being highly ambiguous, 
is not allowed because it changes the topology. 

In the context of the deconfinement transition this problem so far has not 
been addressed. In \cite{e96} it has been noted that the considerations there
could not be extended to the confining phase because of the occurrence of 
entangled structures rather than of simple loops. Otherwise the indicated 
problem so far has not been mentioned in the respective literature.

In U(1) gauge theory in four dimensions the related problem has recently
been solved using mappings which preserve homotopy \cite{krw94,krw95}. 
This has lead to an unambiguous characterization of the phases by the
presence or absence of a current network which is topologically nontrival 
in all directions. More generally this means presence or absence of an 
infinite network where the definition of ``infinite'' on the finite lattice 
is dictated by the boundary conditions \cite{krw96}. For periodic boundary 
conditions it has been checked in detail that neither size nor extension 
but precisely the topological properties are the crucial ones.

These results suggest to try an analogous characterization of the
deconfinement transition. In that case the percolation phenomenon of the
monopole currents should occur only in the three space directions. 
Due to the fact that the characterization is by the indicated 
topological properties there is no problem giving a consistent prescription 
also in the three-dimensional subspace.

Within the latter respect a conceptual problem exists for the other quantities
considered so far, i.e. for monopole density \cite{bsw91,ekms95}, sizes of 
monopole loops \cite{ekms95,kms95} and quantities related to cluster size 
\cite{bmm92}. We have checked, using larger lattices and higher statistics, 
that such quantities also empirically do not lead to a convincing 
characterization.

The indicated characterization by the topology of networks has the 
advantage that it is insensitive to short range fluctuations. In fact, as 
has been shown in the case of U(1) gauge theory \cite{krw94,krw95,krw96}, 
and as turns out also here, it is solely sensitive to the percolation 
phenomenon, i.e.~to extreme long distance properties. In practice the 
characteristic probability for the occurrence of a network which is 
nontrivial in the three space directions, taking sharp values 0 and 1, 
allows determination of the phases at very low computational cost.

In addition to showing that the topological properties signal the percolation
transition it is to be checked wether its transition point coincides with that 
of the deconfining transition of the fields. This is important because 
there are examples where there is no such coincidence \cite{e72,abl87,bfk94}. 
In the only work \cite{bmm92} in which percolation is discussed in the context 
of the deconfinement transition, the coincidence of the respective transition
points is considered to be an open question.

It has been observed in Ref.~\cite{bmm92} that loops of monopole currents 
being nontrivial in the time direction survive in the deconfinement phase. 
Such loops have also been studied in Ref.~\cite{e96} by using the concept of 
a ``wrapping number''. This quantity has already been introduced
in \cite{bls94}. It has been shown, because of current conservation, to 
equal the net current flow \cite{krw94}. In U(1) gauge theory it is known to 
fail to characterize the transition \cite{bls94,krw94}. Thus it appears
desirable to investigate in more detail what happens in time direction which, 
as mentioned above, should not be involved in the percolation. 

A further question is wether the nontrivial loops in time direction are
really simple loops and not, in general, networks with a similar number
of contacts (crossings of monopole currents) as have been observed in U(1) 
theory. The mean numbers per volume of contacts there turn out to be rather 
insensitive to sizes of lattices and of monopole structures \cite{krw94}, 
which indicates a rather uniform density of the contacts within the monopole 
structures. Thus it should be clarified wether the same phemomenon occurs 
also here.

In the present letter we consider SU(2) gauge theory in the maximally abelian 
projection putting particular effort on careful gauge fixing. We apply the 
method of analyzing networks of monopole currents developed before in U(1) 
gauge theory \cite{krw94,krw95} and show that the percolation transition of 
the monopole currents in the three space dimensions precisely coincides with 
the deconfinement transition of the fields. In addition we investigate detailed
features of the structure of the monopole currents in time direction.

\section{Gauge fixing} \setcounter{equation}{0}

\hspace{0.35cm}
The maximally abelian gauge \cite{klsw87} is obtained by performing gauge 
transformations which maximize the quantity  
     \begin{eqnarray}
       R = \sum_{x,\mu} Tr (\sigma_3 U_{\mu,x}
                                       \sigma_3 U^{\dagger}_{\mu,x}) \quad .
       \label{rmap}
     \end{eqnarray}
The conventional procedure is to perform local gauge transformations 
iteratively throughout the lattice until sufficient accuracy is reached.
By introducing an overrelaxation parameter $\omega$ \cite{mo90} the efficiency
of this method can be considerably improved \cite{hkmmos91}. For this either
the fixed choice $\omega=1.7$ \cite{hkmmos92} or choosing stochastically
\cite{dg89} $\omega=1$ (no overrelaxation) in 10\% of cases and $\omega=2$ 
otherwise has turned out to be advantageous.

Applying the overrelaxation methods one may still get stuck at some local 
maximum of (\ref{rmap}). A method suitable to reach the global maximum is 
simulated 
annealing \cite{kgv83}. We find that for the present purpose to use such a 
technique is indeed crucial. The necessity of using simulated annealing
has recently also been pointed out in the context of abelian potentials
\cite{bbmp95}.

Our procedure for simulated annealing uses Metropolis sweeps based on a 
probability distribution $P(R)\sim\exp(\alpha R)$ in which changes of R 
by random local gauge transformations are proposed. After an appropriate 
number of sweeps at each $\alpha$ the value of $\alpha$ has been increased 
by a suitable amount and the simulations continued at the new value. After 
the simulated annealing steps in addition an overrelaxation algorithm has 
been applied. Thus essentially first annealing finds the proper maximum and 
then overrelaxation quickly determines the precise numerical value of its
location. 

To have some control of the quality of the procedure we have always
applied it to several gauge copies (created by random gauge transformations 
of one configuration). We then have considered the data separately for the 
best, the first and the worst copy. By varying the quality of the procedure
we have checked that it is appropriate to associate ``best'' to the largest 
value of $R$ and ``worst'' to the smallest one.

In standard runs we used 20 $\alpha$ values and 10 sweeps at each of them 
for annealing. We stopped the overelaxation steps after $Z=10^{-7}$ has been 
reached, where $Z$ is given by \cite{hkmmos91}
      \begin{eqnarray}
        Z = \frac{1}{L^4} \sum_{x} [(X_{x}^1)^2 + (X_{x}^2)^2] \quad ,
      \end{eqnarray}
with $X_x = \sum_{\mu} ( U_{\mu,x} \sigma_3 U^{\dagger}_{\mu,x} +
U^{\dagger}_{\mu,x-\mu} \sigma_3 U_{\mu,x-\mu})$ , and considered 4 gauge 
copies. 
To check the reliability of the standard run choices we performed high 
quality runs with 600 $\alpha$-values and 100 sweeps at each of them for
annealing. Only minor deviations 
from the best copy results of the standard runs have been observed. These
deviations have remained within the statistical errors of the measured
quantities.

\section{Topological analysis} \setcounter{equation}{0}

\hspace{0.35cm}
In the maximally abelian gauge the coset decomposition of SU(2) with respect 
to the Cartan subgroup U(1) is $U_{\mu,x}=w_{\mu,x}u_{\mu,x}$ 
with $tr(w_{\mu,x}\sigma_3)=0$ and where 
$u_{\mu,x}=\exp(i\theta_{\mu,x}\sigma_3)$ is diagonal. The abelian physical 
flux $\bar{\theta}_{\mu\nu,x}$ is given by
$\theta_{\mu\nu,x}=\bar{\theta}_{\mu\nu,x}+ 2\pi n_{\mu\nu,x}$ with 
$n_{\mu\nu,x} = 0,\pm1,\pm2$ and $\bar{\theta}_{\mu\nu,x}\in[-\pi,\pi)$ 
\cite{dt80}. Defining the monopole currents related to the links of the dual 
lattice by 
  \begin{eqnarray}
     J_{\mu,x} = \frac{1}{2\pi}\epsilon_{\mu\nu\sigma\rho}(
     \bar{\theta}_{\sigma\rho,x+\mu+\nu}-
     \bar{\theta}_{\sigma\rho,x+\mu})
     \label{jdef}
  \end{eqnarray}
the conservation law 
\begin{equation}
\sum_{\mu}(J_{\mu,x}-J_{\mu,x-\mu})=0 
     \label{cc}
\end{equation}
has the simple geometrical meaning that incoming and outgoing currents 
compensate at each site of the dual lattice. 

We define current lines in terms of the currents as follows: for $J_{\mu,x}=0$ 
there is no line on the link, for $J_{\mu,x}=\pm 1$ there is one line, and 
for $J_{\mu,x}=\pm 2$ there are two lines, in positive or negative direction, 
respectively. Networks of currents are connected sets of current lines.
For a network {\bf N} disconnected from the rest the net current flow 
$\vec{f}$ has the components 
\begin{equation}
f_{\mu_3} = \sum_{x_{\mu_0}x_{\mu_1}x_{\mu_2}}J_{\mu_3,x} \quad \mbox{for}
\quad J_{\mu,x}\in \mbox{{\bf N}} \quad .
\end{equation}
By (\ref{cc}) the net current flows of the occurring networks have to sum up 
to zero.

The topological characterization of networks is based on the following
observation. The elements of the fundamental homotopy group are equivalence 
classes of paths starting and ending at a base point $b$ which can be 
deformed continuously into each other. The generators of the group may be 
obtained embedding a sufficiently dense network {\bf N} into its space {\bf X} 
and performing suitable transformations which preserve homotopy. 
If a given network {\bf N} does not wrap around in all directions, then only 
the generators of a subgroup are produced. This fact can be utilized for
an unambiguous characterization of networks.

In practice we choose one vertex point of {\bf N} to be the base point $b$ 
and consider all paths which start and end at $b$. A mapping which shrinks 
one edge to zero length preserves the homotopy of all of these paths. 
Therefore, by a sequence of such mappings we can shift all other vertices to 
$b$ without changing the group content until we finally obtain a bouquet of 
paths which all start and end at $b$. In this procedure we describe a path by 
a vector which is the sum of oriented steps along the path. The bouquet
vectors then form a matrix which is to be analyzed with respect to its
generator content. This is achieved by a modified Gauss elimination procedure
which respects current conservation.

Thus this analysis, on which more details are given in \cite{krw94,krw95},
tells in which directions the network is nontrivial, which are the directions
for which generators are found (i.e.~for which bouquet loops occur which wrap
around the torus).

\section{Results} \setcounter{equation}{0}

\hspace{0.35cm}
Our Monte Carlo simulations have been performed with the Wilson action 
on lattices $8^3\times 4$, $12^3\times 4$ and $18^3\times 4$. The measurements
with gauge fixing and analysis of configurations have been separated by 100 
simulation sweeps. Typically about 100 measurements have been evaluated at
each $\beta$-value.

Figure 1 shows the probability $P_{\mbox{\scriptsize net}}$ for the occurrence 
of a network which is nontrivial in the three space directions as function 
of $\beta$ which we have obtained on lattices $18^3\times 4$, $12^3\times 
4$ and $8^3\times 4$ for best, first and worst gauge copies. The line along
the best copy results is drawn to guide the eye. It is seen that as soon as the 
lattice is large enough one gets a clear signal, $P_{\mbox{\scriptsize net}}$ 
taking the value 1 in the confining phase and 0 in the deconfining phase.
In particular, it turns out that for increasing lattice size the transition
point approaches the one determined from Polyakov loop data (where the most
recent value is $\beta_c=2.29895(10)$ \cite{emsz91}). 

From Figure 1 it can also be seen that going from the data of the best copy 
via the ones of the first copy to those of the worst copy the transition 
appears to shift slightly to the right. The smallness of this shift shows
that the careful gauge fixing described in Sect.~2 is successful. Its
direction shows that a shift due to bad gauge fixing would have the same 
direction as one due to smaller finite size effects on larger lattices. 
Therefore, reliable gauge fixing is crucial for the present purpose. 

With respect to the number of nontrivial networks we find that
in the confining phase there is always (in all of our measurements) just one 
nontrivial network and correspondingly its net current flow is zero. In the 
deconfinement phase there is no network which is nontrivial in the three space 
directions. 

The percolation phenomenon in three dimensions observed here is seen to be
completely analogous to that found in U(1) gauge theory in four dimensions.
Similarly as in the U(1) case various alternative possibilities have been 
checked. It again has turned out that precisely the indicated 
topological properties are the crucial ones and give a completely 
consistent description. 

Next we consider the structure of the monopole currents in time direction. 
The probability 
for finding a current network which is nontrivial in time direction in the
$\beta$ range considered is one on lattices $18^3\times 4$ and $12^3\times
4$ , i.e. in these cases we find always such a network. On the $8^3\times 4$ 
lattice it is one at $\beta=2.2$ and decreases for larger $\beta$ , at
$\beta=2.4$ taking a value of about 0.6 for the best gauge copy and of about
0.8 for the worst copy.

Thus it turns out that nontrivial topology in time direction is certain in both 
phases as soon as the lattice is large enough. Again it becomes obvious that 
the $8^3\times 4$ lattice is too small to produce typical results. From the
comparison of best and worst copies it is again seen that worsening gauge 
fixing causes a change in the same direction as one due to smaller finite 
size effects on larger lattices. Thus careful gauge fixing is important
also here.

A particular feature is that in time direction quite a number of nontrivial
networks occur in the deconfinement phase. Figure 2 gives the average numbers 
per lattice volume of networks being nontrivial in time direction and of simple
nontrivial loops in this direction as function of $\beta$ , which we have 
obtained on lattices $18^3\times 4$, $12^3\times 4$ and $8^3\times 4$ for best 
gauge copies. Obviously there is little dependence on the lattice size.

 From Figure 2 it is seen that though there is some fraction of simple loops 
in general one has to deal with networks. The observation of a particular
fraction at a given value of $\beta$ is related to the observation
already made in U(1) lattice gauge theory \cite{krw94} that the average
number of contacts of current lines per size has a definite value which is 
rather independent of the sizes of the lattice and of the particular current 
structures. 

The number of contacts at a site is defined by the number of 
lines arriving at the site (or, equivalently, departing from it) minus one. 
Figure 3a shows the average number of contacts of current lines per size 
$n_{\mbox{\scriptsize c}}$ as function of $\beta$ for all networks and 
Figure 3b $n_{\mbox{\scriptsize c}}$ for those nontrivial in time direction. 
It is seen that for the latter ones one gets only slightly larger values. 

Thus the situation is analogous as in U(1) gauge theory where the numbers 
for larger networks are only slightly larger than the overall ones 
\cite{krw94}. The explanation is that the density of contacts gets uniform
as soon as the structures are sufficiently extended. To demonstrate this
effect in the present context we show in Figure 3c that a slight increase 
is also observed if the plaquette-like current loops are omitted. 
It is seen that for the largest lattice then about the same values 
as in Figure 3b are reached.

The number of contacts turns out to decrease with $\beta$ as one should expect. 
The numerical value at the transition point is about a factor 1.5 smaller
as in U(1) theory, i.e.~the numbers $n_{\mbox{\scriptsize c}}$ are of the same 
order of magnitude. Thus the observed current networks are of the same type.

If there is more than one nontrivial network, nonzero net current flows become 
possible \cite{krw94}. In the case under consideration they, in fact, occur. 
Figure 4 shows the average numbers per volume of networks with net current 
flows of modulus 0, 1 and 2 in time direction as function of $\beta$ which 
we have got on lattices $18^3\times 4$, $12^3\times 4$ and $8^3\times 4$ for 
the best gauge copy. Apparently $|f_0|=1$ is most frequent while $|f_0|=2$ 
is relativly rare. Higher values have not been observed. Again it is seen 
that there is little dependence on the lattice size.

Possibilities to explain the features occurring in the deconfinement phase 
by 't Hooft-Polyakov monopoles have been discussed in various papers. The
situation has been reviewed by Smit and van der Sijs \cite{ss91} who have 
presented detailed analytical investigations. The picture developed by these
authors, in fact, predicts nontrivial current loops in time direction to
persist in the deconfinement phase. Some caution with this picture, based 
on classical solutions and dimensional reduction arguments, because of the 
number of assumptions involved appears, however, appropriate.

A simpler explanation may focus on the fact that the extension in time 
direction is really very small. Thus, since the current distribution should 
be sufficiently uniform, one can generally expect easy bridging of the torus 
by monopole currents in that direction.

\section*{Acknowledgments}

\hspace{3mm}
We would like to thank A.~Weber for helpful conversations. One of us (W.K.)
wishes to thank M.~M\"uller-Preussker and his group for their kind
hospitality.
This research was supported in part under DFG grant 250/13-1.

\newpage
\renewcommand{\baselinestretch}{1.6}
\small\normalsize

\section*{Figure captions}

\begin{tabular}{rl} 

Fig.~1. & Probability $P_{\mbox{\scriptsize net}}$ as function of $\beta$\\
   & on lattices $18^3\times 4$, $12^3\times 4$ and $8^3\times 4$ \\
   & for best (circles), first (squares) and worst (diamonds) gauge copies.\\

Fig.~2. & Numbers per volume $n_{\mbox{\scriptsize s}}$ of networks nontrivial 
	  in time direction (circles) \\&  and 
          of simple loops nontrivial in time direction (squares) as function \\
        & of $\beta$ on lattices $18^3\times 4$, $12^3\times 4$ and 
          $8^3\times 4$ for best gauge copy.\\

Fig.~3. & Average number per size $n_{\mbox{\scriptsize c}}$ of contacts of 
	  current lines, (a) for all networks, \\& (b) for networks nontrivial
          in time direction, (c) for networks except \\& plaquet-type ones,
         on lattices $18^3\times 4$ (circles), $12^3\times 4$ (squares)
          \\& and $8^3\times 4$ (diamonds) for best gauge copy.\\

Fig.~4. & Numbers per volume $n_{\mbox{\scriptsize f}}$ of networks with 
	  flows $|f_0|=0$ 
          (circles), $|f_0|=1$ \\& (squares) and 
         $|f_0|=2$ (diamonds) in time direction as function of $\beta$ \\
        & on lattices $18^3\times 4$, $12^3\times 4$ and $8^3\times 4$ 
         for best gauge copy.\\

\end{tabular}

\newpage

\begin{figure}[tb]
%\centering
\caption{}
\includegraphics{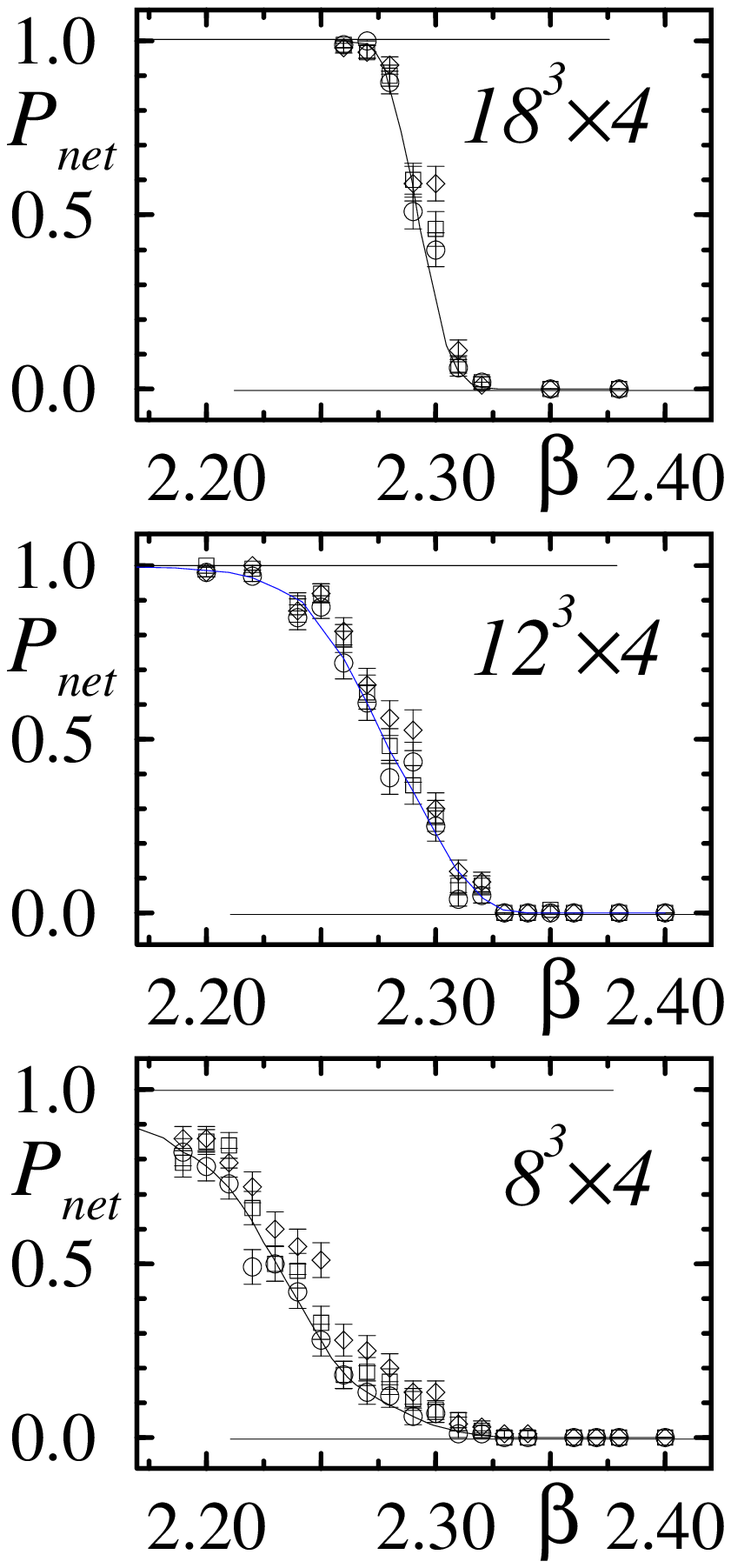}
\end{figure}

\begin{figure}[tb]
\centering
\caption{}
\includegraphics{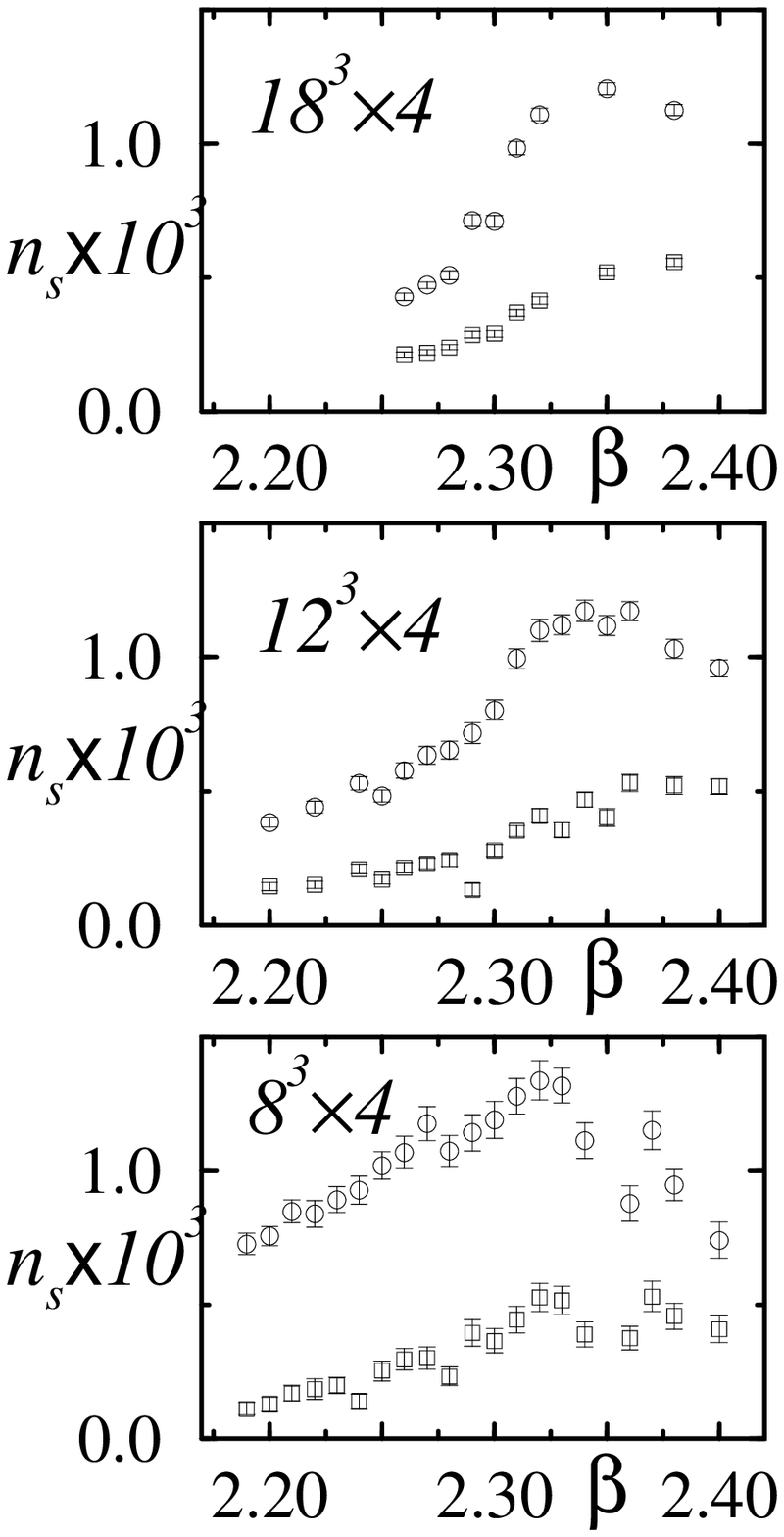}
\end{figure}

\begin{figure}[tb]
\centering
\caption{}
\includegraphics{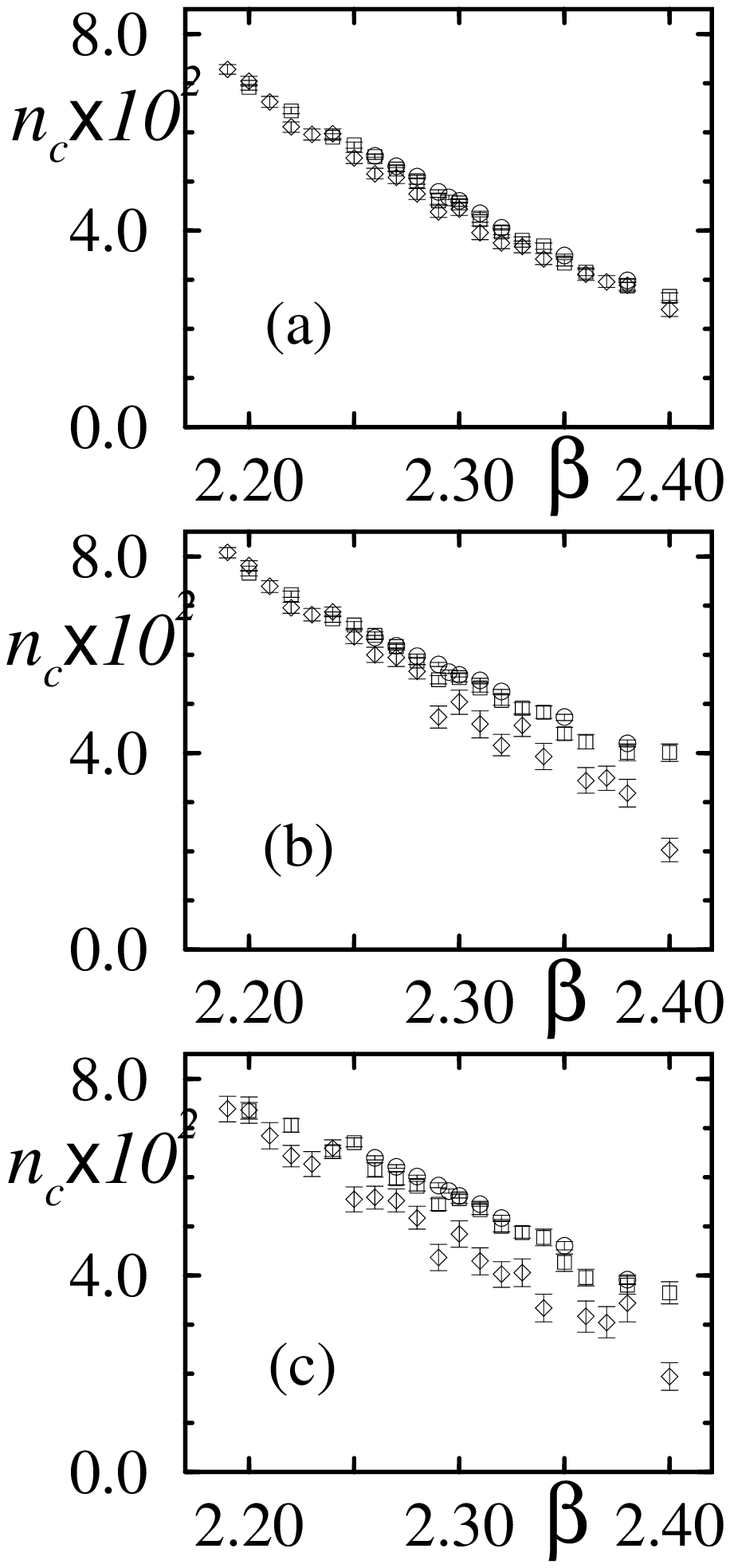}
\end{figure}

\begin{figure}[tb]
\centering
\caption{}
\includegraphics{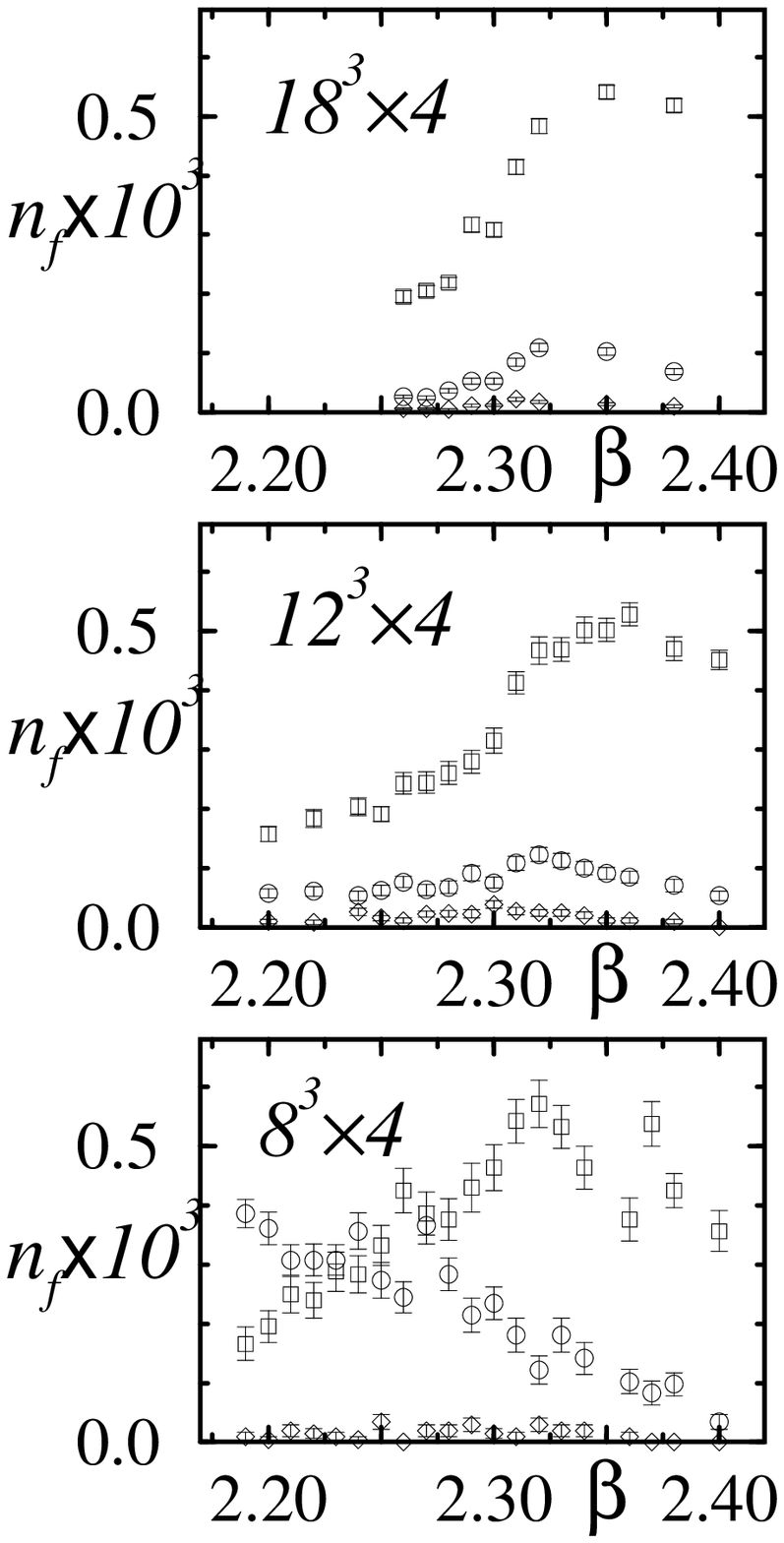}
\end{figure}

\end{document}